\documentclass[prl,twocolumn,showpacs,amsmath,amssymb,superscriptaddress]{revtex4}
\usepackage{dcolumn}
\usepackage{bm}
\usepackage{natbib}
\usepackage{amssymb}
\usepackage{dcolumn}
\usepackage[dvips]{graphicx}

\def\gamnas{Ga$_{1-x}$Mn$_x$As}

\def\tc{T${_{\mathrm{c}}}$}
\def\xx{_{\mathrm{xx}}}
\def\xy{_{\mathrm{xy}}}
\def\f{_{\mathrm{F}}}
\def\k{_{\mathrm{K}}}
\begin{document}

\title{Electronic structure of ferromagnetic semiconductor \gamnas\ probed by sub-gap magneto-optical spectroscopy}

\author{G. Acbas}
\author{M.-H. Kim}
\affiliation{Department of Physics, University at Buffalo, SUNY, Buffalo, New York 14260, USA}
\author{M. Cukr}
\author{V. Nov\'ak}
\affiliation{Institute of Physics ASCR, v.v.i., Cukrovarnick 10, 162 53 Praha 6, Czech Republic}
\author{M. A. Scarpulla}
\affiliation{Department of Materials Science and Engineering and Lawrence Berkeley National Laboratory, University of California,
Berkeley, California 94720, USA}
\author{O. D. Dubon}
\affiliation{Department of Materials Science and Engineering and Lawrence Berkeley National Laboratory, University of California,
Berkeley, California 94720, USA}
\author{T. Jungwirth}
\affiliation{Institute of Physics ASCR, v.v.i., Cukrovarnick 10, 162 53 Praha 6, Czech Republic}
\affiliation{School of Physics and Astronomy, University of Nottingham, Nottingham NG7 2RD, United Kingdom}
\author{Jairo Sinova}
\affiliation{Department of Physics, Texas A M University, College Station, Texas 77843-4242, USA}
\affiliation{Institute of Physics ASCR, v.v.i., Cukrovarnick 10, 162 53 Praha 6, Czech Republic}
\author{J. Cerne}
\affiliation{Department of Physics, University at Buffalo, SUNY, Buffalo, New York 14260, USA}


\begin{abstract}
We employ Faraday and Kerr effect spectroscopy in the infrared range to investigate the electronic structure of \gamnas\ near the Fermi energy. The band structure of this archetypical dilute-moment ferromagnetic semiconductor has been a matter of controversy, fueled partly by previous measurements of the unpolarized infrared absorption and their phenomenological impurity-band interpretation. The infrared magneto-optical effects we study arise directly from the spin-splitting of the carrier bands and their chiral asymmetry due to spin-orbit coupling. Unlike the unpolarized absorption, they are intimately  related to ferromagnetism and their interpretation is much more microscopically constrained in terms of the orbital character of the relevant band states. We show that the conventional theory of the disordered valence band with dominant As $p$-orbital character and coupled by kinetic-exchange to Mn local moments accounts semi-quantitatively for the overall characteristics of the measured infrared magneto-optical spectra.
\end{abstract}

\date{\today}
\maketitle

The study of diluted magnetic semiconductors (DMSs) based on III-V compounds, in particular \gamnas,
has been a very active area of research
over the past two decades ever since the discovery of carrier mediated ferromangnetism in these materials \cite{Ohno:1992_a}.
\gamnas\ can exhibit metallic as well as insulating behavior depending on the doping and growth procedures, with
the onset of ferromagnetism occurring near the insulator-to-metal transition. The perception that this transition occurs in a disordered valence band as in conventional p-doped semiconductors \cite{Dietl:1997_a,Ohno:1998_a,Jungwirth:1999_a,Dietl:2000_a} and that many of the key magnetic properties are reminiscent of common itinerant ferromagnets \cite{Jungwirth:2006_a,Novak:2008_a} makes \gamnas\ an archetypical ferromagnetic semiconductor system. The notion has inspired numerous studies that contributed to our understanding of basic magnetic and magneto-transport properties of DMSs and led to discoveries of new  effects and device concepts, including various types of spintronic transistors \cite{Dietl:2008_b}.
Nevertheless, the character of states near the Fermi energy in ferromagnetic \gamnas\ is still a matter of controversy \cite{Burch:2006_a,Jungwirth:2007_a,Rokhinson:2007_a,Garate:2008_c,Stone:2008_a,Tang:2008_a,Ando:2008_a} fueled in part by phenomenological
interpretations of optical spectroscopy studies \cite{Burch:2006_a,Tang:2008_a,Ando:2008_a} which are inconsistent with the conventional semiconductor valence-band picture of the material.

A broad peak around $\sim 200$ meV observed in unpolarized infrared absorption spectra has been interpreted by placing the Fermi energy in a narrow impurity band detached from the semiconductor valence band, which persists  on the metallic side of the transition \cite{Burch:2006_a}. Competing with this phenomenological picture, a semi-quantitative agreement with the experimental peak-position and magnitude of the absorption has been obtained from microscopic calculations based on the disordered valence-band theory and considering transitions within the valence band \cite{Jungwirth:2007_a}. The merit of the infrared spectroscopy is that it probes electronic states near the Fermi energy. The unpolarized absorption spectra, however, do not distinguish between transitions originating from states corresponding to intrinsic \gamnas\ and from extrinsic  impurities, and do not provide specific information on the magnetic and orbital character of states involved in the optical transitions. A quantitative microscopic modeling of these spectra is complicated by the relatively strong intentional and unintentional disorder in these heavily doped semiconductors and by the vicinity of the metal-insulator transition \cite{Jungwirth:2007_a}. Without microscopic modeling, the absence of a direct link between the measured spectra and the exchange-split and spin-orbit coupled electronic states leaves a relatively unconstrained space for inferring   phenomenological pictures of the underlying band structure.

In this paper we present results of Faraday and Kerr measurements in the mid and near-infrared
spectral range (115-1300 meV) for several ferromagnetic \gamnas\ samples grown by low temperature molecular beam epitaxy (LT-MBE) \cite{Holy:2006_a} and a sample
grown by a combination of ion implantation and laser pulsed melting (II-LPM) \cite{Scarpulla:2003_a,Scarpulla:2008_a}. Our work complements previously reported higher energy (starting from 600~meV) magneto-optical measurements in \gamnas\ \cite{Ando:1998_a,Kuroiwa:1998_a,Beschoten:1999_a,Szczytko:1999_a,Lang:2005_a,Chakarvorty:2007_a,Ando:2008_a,Komori:2003_a}(in the thin film limit the magnetic circular dichroism is approximatively twice the Faraday ellipticity). The interpretation of these higher energy  spectra is complicated by the fact that they are sensitive to many desired as well as undesired transitions near the host semiconductor absorption edge. The infrared magneto-optical data studied here provide a unique probe into the nature of electronic states near the Fermi level. In contrast to the unpolarized absorption which is proportional to the longitudinal conductivity, the Faraday and Kerr effects arise from the transverse conductivity component. They represent a finite frequency analogue of the anomalous Hall effect with which they share the common exchange-splitting and spin-orbit coupling origin. The anomalous Hall effect has been described microscopically on a semi-quantitative level using the conventional semiconductor valence band theory \cite{Jungwirth:2003_b}. Here we show that by extending the same microscopic theory to finite frequencies we capture key characteristics of the measured infrared magneto-optical spectra, again on a semi-quantitative level, as predicted by early theory studies \cite{Sinova:2003_a,Hankiewicz:2004_a}. We have not performed alternative calculations that would correspond to any of the phenomenological variants of the impurity-band model. For calculating the transverse conductivity one has to specify the microscopic spin and orbital character of the relevant states. As explained in detail elsewhere \cite{Masek:2009_tbp}, we have not found any perceivable microscopic representation of the impurity band models that yields a detached impurity band in ferromagnetic \gamnas\ materials.

\begin{figure}
\includegraphics[width=.84\columnwidth]{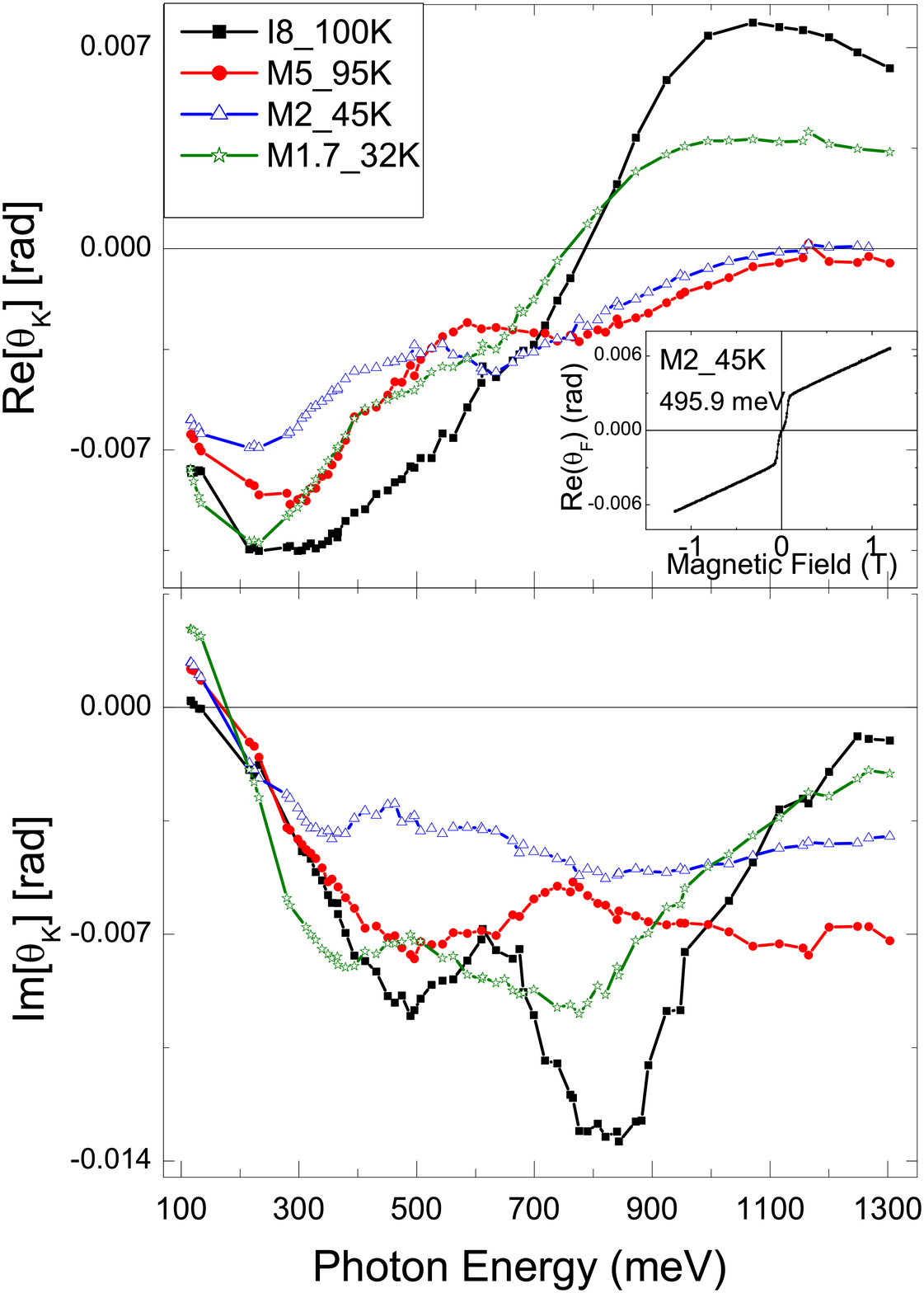}
\caption{ (Color on-line) Measured Kerr rotation (top) and  ellipticity (bottom) vs. photon energy. Top insert: Re$(\theta\f)$ vs magnetic field for sample M2\_45K measured at photon energy of 495.9 meV and $T=10K$}
\label{kerr}
\end{figure}

The measurements were performed with a polarization modulation technique using various
gas and semiconductor laser sources, as well as a custom-modified double pass prism monochromator
with a Xe  light source. For details of the experimental technique see Refs. \cite{Cerne:2003_a,Kim:2007_a,Kim:2009_a}.
The method was tested by studying semiconductor materials with known behavior in mid-infrared region \cite{Kim:2009_a}.
The magnetic field dependence of the complex Faraday $\theta\f$ and Kerr $\theta\k$ angles is measured at each wavelength and
various temperatures from 6K to 300K. At temperatures below \tc\, the magnetic field dependence of the
rotation Re$(\theta)$ and ellipticity Im$(\theta)$ exhibits a saturation behavior
due to alignment of the magnetization with the external field perpendicular to the sample surface. The behavior is superimposed on a linear dependence at higher field due to the response of the wedged GaAs substrate,
the \gamnas\ film, and the cryostat windows (see insert in Fig.~\ref{kerr} ). We subtract the linear component and record the rotation and ellipticity signals corresponding to saturation magnetization.
In principle, this may not completely remove the paramagnetic contribution from the signal,
but this signal is an order of magnitude weaker than the ferromagnetic contribution in this spectral range
and is generally represented by a Brillouin function of the external magnetic field and temperature.
For our measurements (at temperatures above 10 K and magnetic fields below 1.5 T), this dependence is linear in magnetic field.
Since the Faraday (Kerr) signal \cite{Cerne:2003_a,Kim:2007_a} is normalized by the total
intensity of the transmitted (reflected)  light, strong spectral features in the transmittance (reflectance)
of the paramagnetic regions could produce artifacts in the Faraday (Kerr) signal.
We may expect the contribution to the total transmission (reflection) of light through (from) paramagnetic regions to slightly shift the absolute value of  $\theta\f$ ($\theta\k$). However, since the transmittance (reflectance)
varies slowly in the mid and near-infrared range, as confirmed by measurements, these spectral artifacts in $\theta\f$ ($\theta\k$) are expected to play a negligible role.

\begin{figure}
\includegraphics[width=.84\columnwidth]{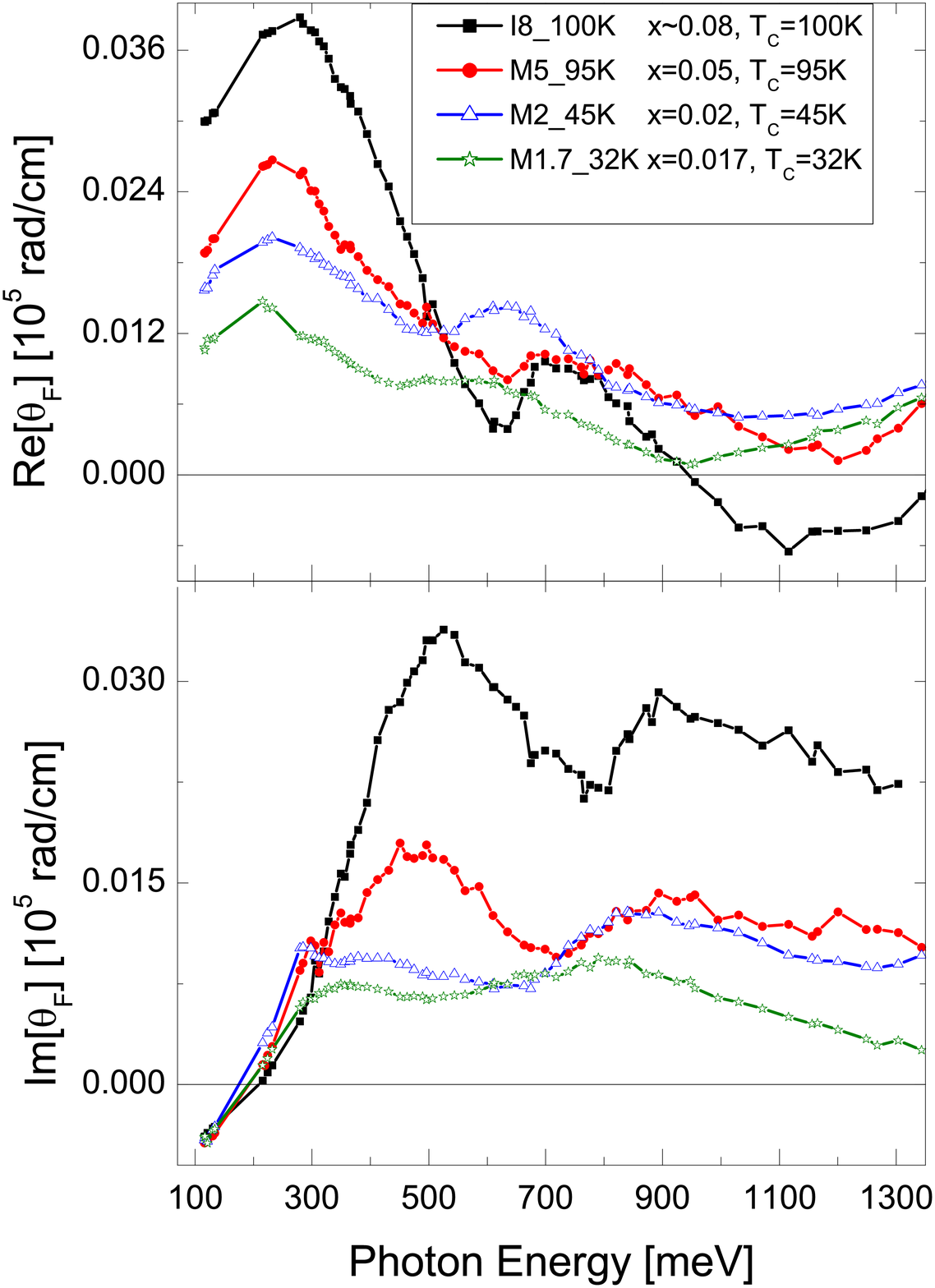}
\caption{ (Color on-line) Measured Faraday rotation (top) and  ellipticity (bottom) vs. photon energy.}
\label{faraday}
\end{figure}

\begin{figure}[h]
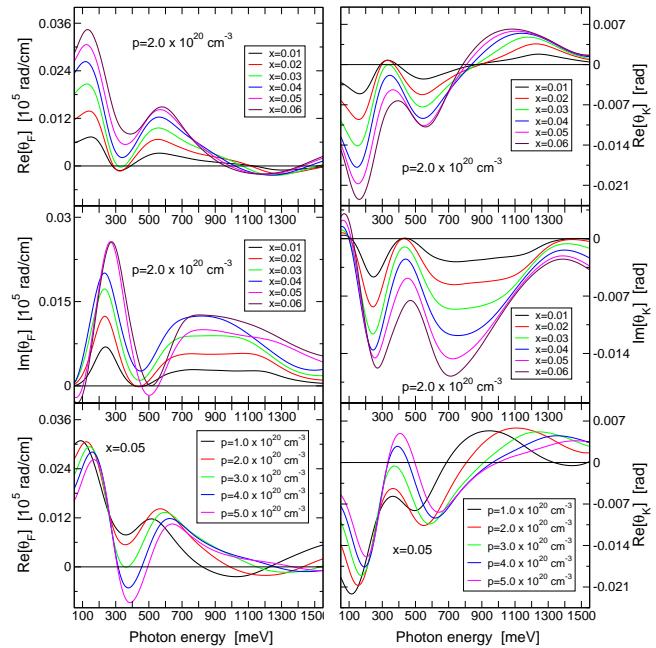

\vspace*{0cm}
\includegraphics[width=0.49\columnwidth]{theta_F_all.eps}
\includegraphics[width=0.482\columnwidth]{theta_K_all_rl.eps}
\vspace*{0cm}
\caption{ (Color on-line) Calculated Faraday rotation Re$(\theta\f)$ and ellipticity Im$(\theta\f)$ (left  panels)  and
Kerr rotation Re$(\theta_K)$ and ellipticity Im$(\theta_K)$  (right  panels) vs. photon energy for samples
of 50nm thickness.
Upper and middle panels show the rotation and ellipticity for several Mn$_{\rm Ga}$ dopings at fixed hole density respectively.
The bottom panels show the rotation for several hole densities at fixed Mn$_{\rm Ga}$ doping.
}
\label{theory}
\end{figure}

We have studied four \gamnas\ films with different ferromagnetic properties.
The sample M5\_95K was grown by LT-MBE on a GaAs substrate. The \gamnas\ film is 50~nm thick with a uniform nominal Mn doping of 5\%, a Curie temperature $T_c=95$~K, and a hole density $p\sim 2\times 10^{20}$~cm$^{-3}$  estimated from transport measurements.
Sample M2\_45K consists of a 50~nm thick \gamnas\ film with nominal Mn doping of 2\%, $T_c=45$~K and $p\sim 1.5\times 10^{20}$~cm$^{-3}$, and the third LT-MBE sample M1.7\_32K has a 100~nm thick \gamnas\ with nominal Mn doping of 1.7\%, $T_c=32$~K, and $p\sim 1\times 10^{20}$~cm$^{-3}$. Sample I8\_100K was grown by II-LPM and has $T_c\approx 100$~K and a Gaussian-like distribution of Mn ions
with an estimated mean concentration of Mn of $\sim 8$\% and hole density $p\sim 4\times 10^{20}$~cm$^{-3}$.

The spectral dependence of the Faraday and Kerr rotation [Re$(\theta\f)$ and Re$(\theta\k)$] and
ellipticity [Im$(\theta\f)$and Im$(\theta\k)$] taken at 10 K are shown in Figs.~\ref{faraday} and~\ref{kerr}.
All samples present a similar resonant behavior. Unlike the infrared absorption which is proportional to the longitudinal conductivity $\sigma\xx$ \cite{Burch:2006_a}, $\theta\f$ and $\theta\k$
follow the behavior of the off-diagonal component $\sigma\xy$, thus exhibit relatively sharp ($\sim$150 meV) spectral features and sign changes.
The magnitude of the Faraday rotation and ellipticity grows with  Mn concentration.
The rotation spectra are characterized by two extrema separated by about 400~meV and a dip or sign change as
the band gap energy is approached. The ellipticity changes sign in the 200~meV region and has two maxima at higher energy.
We note that features in the rotation are correlated to features in the ellipticity due to the Kramers-Kronig relations.

In the following we compare the measurements with predictions of the theory in which holes reside in an As $p$-orbital dominated valence band which is coupled to Mn $d$-orbital local moments by an antiferromagnetic kinetic-exchange interaction \cite{Jungwirth:2006_a}. The coupling is treated within the mean-field theory and disorder is introduced  through a finite life-time broadening of the valence-band states \cite{Sinova:2003_a,Hankiewicz:2004_a}.
We use the eight-band ${\bf k}\cdot{\bf p}$ model which takes into account the complex band structure of the valence band
in the presence of the kinetic-exchange field and spin-orbit coupling, as well as valence-conduction band transitions. The relatively large carrier doping also leads to band renormalization effects \cite{Berggren:1981_a,Sernelius:1986_a,Zhang:2005_a}
which can modify the band gap and inter-valence band splittings. Whereas electron-electron interactions
contributions from the exchange and correlation effects tend to shift rigidly the valence  and conduction bands towards each other, the impurity scattering from the randomly distributed Mn acceptors leads to a more strongly momentum dependent shift which renormalizes the inter-valence band splittings \cite{Sernelius:1986_a,Zhang:2005_a}.
We have incorporated this in the calculations by shifting rigidly the majority bands only (largest $k_F$) by 10$\times$ (p[$10^{18}$cm$^{-3}])^{1/3}$)~meV \cite{Sernelius:1986_a,Zhang:2005_a}.

We use the calculated complex conductivities $\sigma\xx$ and $\sigma\xy$
in the transmission and reflection formulas for our geometry (thin film on a wedged substrate with multiple scattering effects taken into account) \cite{Kim:2007_a} to determine the complex $\theta\f$ and $\theta\k$.
Due to the low $\sigma\xx$ values in these samples, $\theta\f$ and
$\theta\k$ practically follow $\sigma\xy$ and $-\sigma\xy$, respectively.

In Fig.~\ref{theory} we plot the Kerr and Faraday rotation (top panels) and ellipticity (middle panels)
as a function of the photon energy for a fixed density of carriers $p=2.0\times 10^{20}$ cm$^{-3}$
and  varying Mn$_{\rm Ga}$ doping. For  a fixed Mn$_{\rm Ga}$  doping of 5\% and varying
hole concentration (bottom panel) we only show the Faraday and Kerr rotation. The strong disorder present in the experimental films
is taken into account through a finite life-time of 100~meV \cite{Zhang:2005_a,Jungwirth:2006_a}.
The Faraday rotation has an extremum in the 150-200~meV while the ellipticity changes sign in this region.
These spectral features are due to the transitions from the light-hole band to the heavy-hole band.
Contributions from the split-off band to the light-hole band and the split-off band to the heavy-hole band dominate the Kerr and
Faraday spectra at energies higher than 500-650 meV. Our calculations capture the measured order of magnitude of the
Faraday rotation angle, the separation of the two maxima, and the increase of the angle with increasing  Mn$_{\rm Ga}$ concentration.
Variations in the carrier concentration do not affect the magnitude of the Faraday angle as strongly (see lower left panel in Fig.~\ref{theory}).
The Faraday angle calculations also capture the dip (and a possible change in
sign) of the measured Re$(\theta\f)$ as the energy approaches the band-gap.
Kerr angle calculations (right panels in Fig.~\ref{theory}) show a similar general agreement
with the correct location of the sign change, correct order of magnitude and a separation of the two main peaks in the negative region.

Because of the complexity of the studied materials and approximations employed in the calculations, only a semi-quantitative comparison between theory and experiment is meaningful in case of DMSs.
On this level, the overall agreement of our theoretical infrared magneto-optical spectra with experiment provides strong evidence that the conventional disordered valance band theory of ferromagnetic \gamnas\ captures correctly the magnetic and orbital nature and density of states near the Fermi energy.  We emphasize that our microscopic calculations use no adjustable parameters. This applies to the basic  spin-splitting and orbital characteristics of the bands which we have accounted for within the ${\bf k}\cdot{\bf p}$ model antiferromangetically coupled to the localized Mn d-electrons and whose similar description can be obtained using $spd$-tight-binding model or LDA+U full potential density-functional calculations \cite{Jungwirth:2006_a,Masek:2009_tbp}. It also applies to the additional effects of disorder and interactions on the imaginary and real parts of the hole Green's function; the former has been estimated based on numerical Fermi's golden rule calculations in \gamnas\ \cite{Jungwirth:2006_a} and the latter from detailed band-renormalization studies  in heavily p-doped semiconductors \cite{Sernelius:1986_a,Zhang:2005_a}.

In conclusion, we have presented results of Faraday and Kerr effect measurements in the sub-gap region in order to study the electronic structure of \gamnas. By measuring $\theta\f$ and $\theta\k$, which arise from transitions between exchange-split and spin-orbit coupled states, and after subtracting the paramagnetic background we obtained spectral features corresponding to the ferromagnetically coupled electronic states near the Fermi energy.  Microscopic calculations based on the conventional semiconductor valence band picture of \gamnas\  account for the position of the main spectral features and for the Mn-concentration dependent magnitude of the measured magneto-optical signals.

\acknowledgments
 This work was supported by Research Corporation Cottrell Scholar Award (JC and JS), by
NSF-CAREER-DMR0449899 (JC), by NSF-DMR-0554796,  by DOE Contract No. DE-AC03-76SF00098 (ODD),
by ONR-N000140610122(JS and TJ), by NSF-CAREER--DMR-0547875 (JS), by the SWAN-NRI(JS and TJ),  by
EU Grants FP7-215368 SemiSpinNet and FP7-214499 NAMASTE and from Czech Republic Grants FON/06/E001,
FON/06/E002, AV0Z10100521, KAN400100652, LC510, and Preamium Academiae (TJ and VN).


\end{document}